\documentclass[aps,pra,twocolumn,showpacs,groupedaddress,amsmath,amssymb]{revtex4}
\usepackage{graphicx}% Include figure files
\usepackage{color}
\usepackage{bm}
\usepackage{doi}
\usepackage{float}

\begin{document}
\draft
\def\ds{\displaystyle} 
\title{Topological states in non-Hermitian two-dimensional Su-Schrieffer-Heeger model }
\author{C. Yuce$^1$}
\email{cyuce@eskisehir.edu.tr}
\author{H. Ramezani$^2$}
\affiliation{$^1$Department of Physics, Eskisehir Technical University, Eskisehir, Turkey\\
	$^2$Department of Physics and Astronomy, University of Texas Rio Grande Valley, Brownsville, TX 78520, USA}

\date{\today}
%\pacs{ }
%\keywords{Suggested keywords}
\begin{abstract}
A non-Hermitian topological insulator with real spectrum is interesting in the theory of non-Hermitian extension of topological systems. We find an experimentally realizable example of a two dimensional non-Hermitian topological insulator with real spectrum. We consider two-dimensional Su-Schrieffer-Heeger (SSH) model with gain and loss. We introduce non-Hermitian polarization vector to explore topological phase and show that topological edge states in the band gap exist in the system.
\end{abstract}
\maketitle

\section{Introduction}

The past few years have witnessed a growing amount of interest in the theory of non-Hermitian extension of topological insulators \cite{1d5,ghatakdas}. A Combination of long-lived quantum states immune to decoherence and the unique features of non-Hermitian systems make non-Hermitian topological systems a promising platform for future applications in quantum technology. One can not directly generalize the standard theory of topological insulators to non-Hermitian systems since they have not always real eigenvalues and orthogonal eigenstates. Furthermore, exceptional points (EPs) where at least two eigenstates coalesces can appear only in non-Hermitian systems \cite{kato} and lead to new topological features. This new field of study is still in its infancy and non-Hermitian topological phases have been mainly investigated in one dimensional (1D) topological systems such as Su-Schrieffer-Heeger (SSH) lattices \cite{1d1,1d2,1d3,1d3ekl,1d4,1d6,1d7,1d8,1d9,1d10,1d11,1d12,1d13,1d14,1d15,winding1}, Aubry-Andre chain \cite{aah1} and Kitaev model \cite{kita1,kita2,kita3,kita4,kita5}, in which non-Hermiticity is obtained by introducing asymmetric tunneling amplitudes and/or gain-loss. 

The standard bulk-boundary correspondence doesn't work in non-Hermitian systems \cite{bulkboun01,bulkboun02,bulkboun03,bulkboun04,bulkboun05,bulkboun06,bulkboun07,bulkboun08da}. The so-called non-Hermitian skin effect arises when the topological lattice has asymmetric tunneling amplitudes. In this case, bulk states as well as topological states are localized near edges. On contrary to Hermitian systems, topological phase transition point can not be determined by the translationally invariant form of the non-Hermitian Hamiltonian. We stress that topological phase transition can also be induced solely by gain and loss \cite{takata}. Recently, the idea of pseudo topological insulators has been introduced to explain topological edge states for massive SSH Hamiltonian \cite{pseudo}. 

There is a fundamental difference between the Hermitian and non-Hermitian 1D topological insulating systems. In topologically nontrivial Hermitian systems, edge states disappear when the band gap is closed. However, edge states may acquire an imaginary energy gap in non-Hermitian systems. The complex topological state can be used to generate spontaneous topological pumping \cite{yucepump} and topological lasing \cite{merced,merced2}. The existence of complex topological state in non-Hermitian systems has applications in lasing but of special importance and yet a non-trivial task is to find topological edge states with real eigenvalues in non-Hermitian systems. It was theoretically shown in \cite{aah1} that such states exist in a complex extension of the Aubry-Andre model. Topological zero energy edge state in $1D$ photonic lattice was experimentally realized \cite{sondeney1}. However, no non-Hermitian topological system with real spectrum has been found in the literature in two-dimensions (2D) \cite{2d1,2d2,2d3,2d4,2d5,2d6,2d7,2d8,2d9,2d10,2d11,2d12,2d13}. It is tempting to find topological systems with real eigenvalues in 2D. In this paper, we consider the two-dimensional Su-Schrieffer-Heeger (SSH) model with gain and loss, which can be experimentally realized with current technology in photonics. In the absence of gain and loss, a nontrivial topological phase was shown to exist even if the Berry curvature is zero  \cite{temel1,temel2}. In the presence of gain and loss, we show that the system can have real valued energy eigenvalues for all bulk and topological states. We introduce complex polarization vector and discuss topological invariant in our non-Hermitian system.  

\section{2D SSH model}

\begin{figure}[t]
\includegraphics[width=4.2cm]{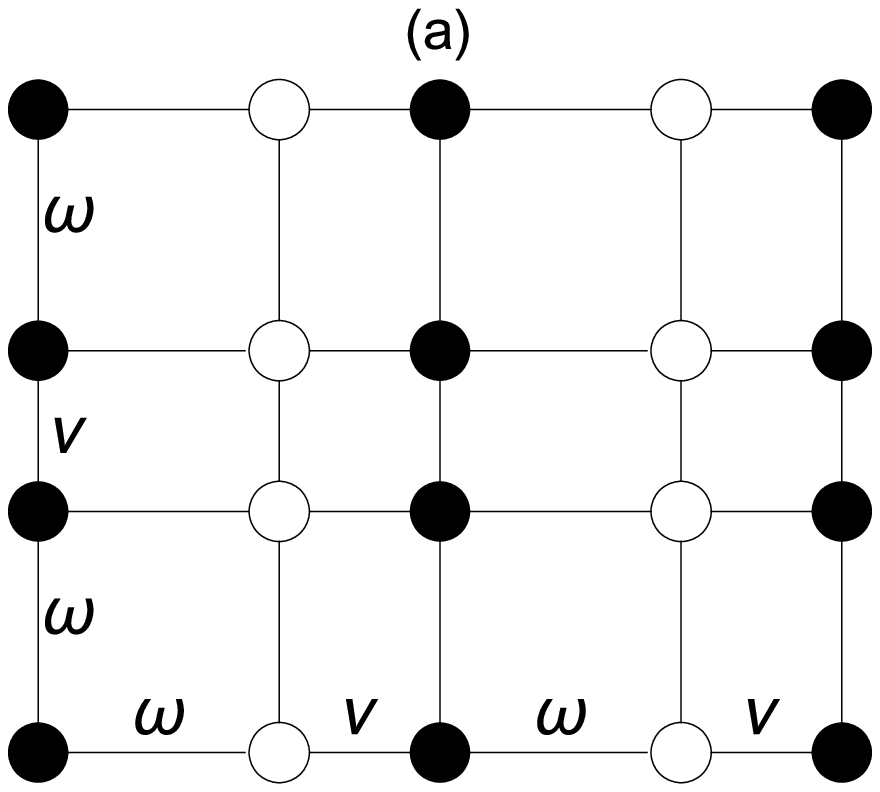}
\includegraphics[width=4.2cm]{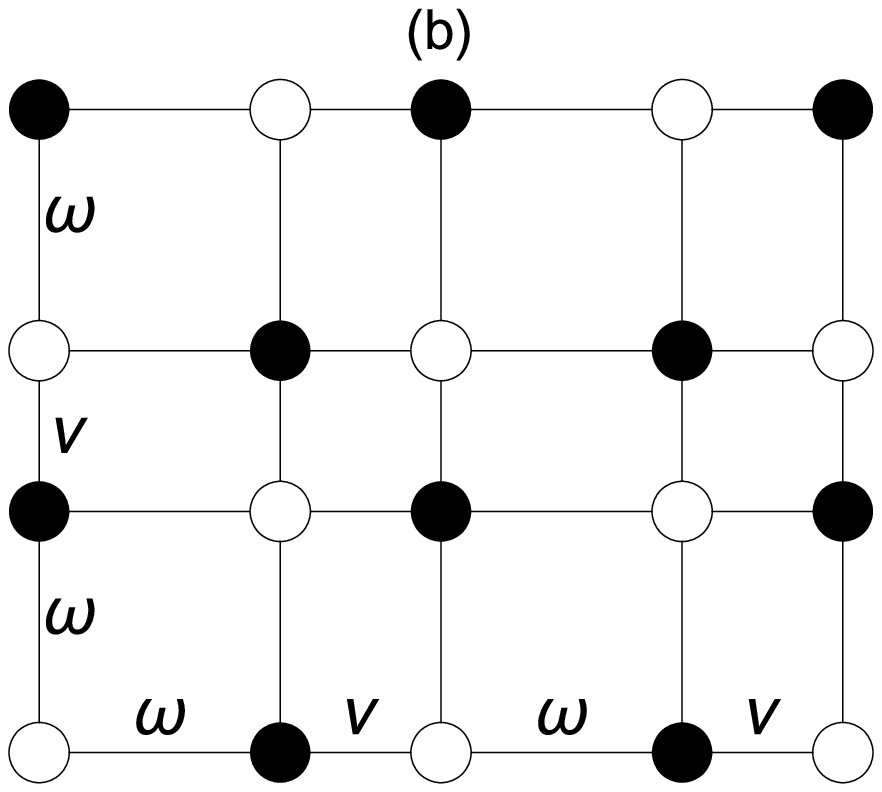}
\caption{  A square lattice with alternating tunneling amplitudes $\omega$ and $\nu$ in each direction. Black circles have gain while unfilled circles have loss. The corresponding non-Hermitian Hamiltonians are given by (\ref{pwdapmu}). The gain and loss impurities alternate only in horizontal ($x$) direction in (a) while they alternate in both horizontal and vertical ($y$) directions in (b). The corresponding energy bands can be real for (a) unless the non-Hermitian degree doesn't exceed a critical number. However, they are complex for (b) for any nonzero value of the non-Hermitian degree.}
\label{fig1}
\end{figure}

\begin{figure}[t]
\includegraphics[width=4.25cm]{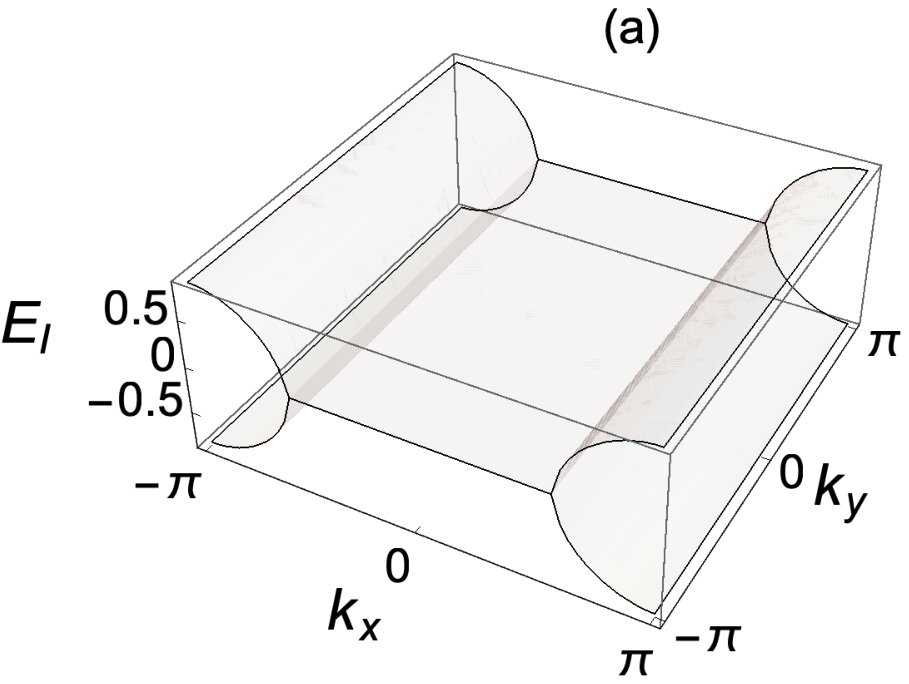}
\includegraphics[width=4.25cm]{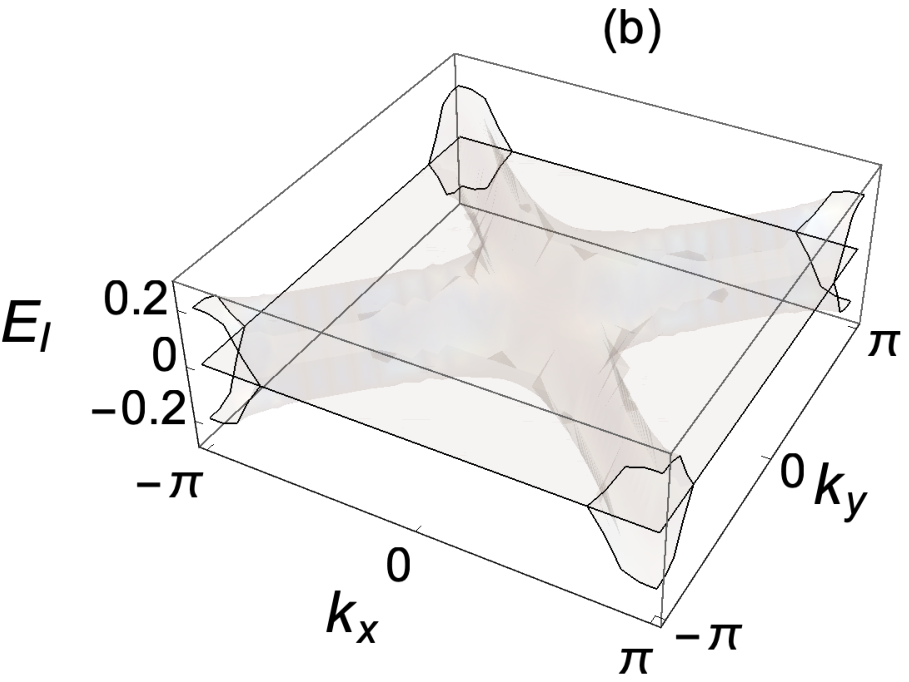}
\caption{ Imaginary part of energy eigenvalue for the system depicted in Fig 1.a  at $\gamma=1$ (a)  and Fig 1.b at $\gamma=0.2$ (b). The tunneling amplitudes are given by $\nu=1$ and $\omega=0.5$. The system in (a) has real spectrum unless $\gamma>|\nu-\omega| $ while the system in (b) has complex spectrum whenever $\gamma\neq 0$.    }
\label{fig2a}
\end{figure}

Consider the 2D Su-Schrieffer-Heeger model with gain and loss, which describes a non-Hermitian square lattice with alternating tunneling amplitudes in each direction. The tunneling amplitudes in horizontal and vertical directions are assumed to be alternating between $\omega$ and $\nu$ as shown in the Fig. \ref{fig1}. There are four sites in the unit cell and one can add gain and loss in the unit cell. The non-Hermitian Hamiltonian we consider is given by
\begin{eqnarray}\label{65748mu}
\mathcal{H}&=&  \sum_{i,j} \Delta_i  ~a_{i+1,j}^{\dagger} ~a_{i,j}+  \Delta_j ~a_{i,j+1}^{\dagger} ~a_{i,j}\nonumber\\
&+&i \sum_{i,j}  {\gamma_{ij}}~a_{i,j}^{\dagger}~ a_{i,j}+H.
 C.
\end{eqnarray}
where  $a_{i,j}^{\dagger} $ and $a_{i,j} $ are the creation and annihilation operators at the site $(i, j)$, respectively, $\ds{\Delta_{i,j}=t +(-1)^{i,j}  ~\delta  }$, $t>0$ is the reference tunneling amplitude and $\delta<t$ is a constant, which are related to the tunneling amplitudes through $\ds{\omega=t-\delta}$ and $\ds{\nu=t+\delta}$ and the real valued parameters $\ds{\gamma_{ij}}$ describe gain/loss strengths at the lattice sites $i,j$. We consider that the system is periodic and hence the gain/loss strengths in the unit cell is given by $\ds{  \{   \gamma_1, \gamma_2, \gamma_3, \gamma_4\}}$. Here we assume that the gain and loss in the system is balanced, which implies $\ds{ \gamma_1+ \gamma_2+\gamma_3+\gamma_4 =0}$.  \\
The gain and loss can be arranged in various ways in the unit lattice. Figure \ref{fig1} depicts two such configurations where black and white circles are for gain and loss impurities. The non-Hermitian strengths in the unit cell are given by $\ds{  \{   \gamma,- \gamma, \gamma,- \gamma\}}$ and $\ds{  \ \{ -  \gamma, \gamma, \gamma,- \gamma\}}$ for (a) and (b), respectively, where $\ds{\gamma}$ is a constant. In both cases, the systems have balanced gain and loss. We stress that the reality of the spectrum depends sensitively on the gain and loss arrangements in the system.\\
Applying the Fourier transformation, we get the matrix form of the Hamiltonian in the $\bf{k}$-space 
\begin{equation}\label{r502uncmu}
\mathcal{H} = \left(\begin{array}{cccc}i\gamma_1&  \omega+\nu e^{-ik_x}  & \omega+\nu e^{-ik_y}  & 0 \\  \omega+\nu e^{ik_x}  & i\gamma_2 & 0 &  \omega+\nu e^{-ik_y}     \\ \omega+\nu e^{ik_y}  &0 & i\gamma_3 &  \omega+\nu e^{-ik_x}        \\0 & \omega+\nu e^{ik_y}  &  \omega+\nu e^{ik_x}  & i\gamma_4\end{array}\right)
\end{equation}
To study symmetry properties of the two choices in Fig. \ref{fig1}, we rewrite the corresponding Hamiltonians using the Pauli spin matrices. If we specify the gain and loss strengths in the unit cell, we get
\begin{eqnarray}\label{pwdapmu}
\mathcal{H}_1 &=&\mathcal{H}_0+i~\gamma~ \mathcal{I}\otimes \sigma_z\nonumber\\
\mathcal{H}_2 &=&\mathcal{H}_0-i~\gamma~ \tau_z\otimes \sigma_z
\end{eqnarray}
where ${ \mathcal{I} }$ is the identity matrix and their Hermitian part $\mathcal{H}_0$ reads
\begin{eqnarray}\label{pdbxmu}
\mathcal{H}_0 &=& (\omega+\nu \cos k_x)~ \mathcal{I} \otimes \sigma_x- \nu \sin k_x ~ \mathcal{I} \otimes \sigma_y\nonumber\\
&+&(\omega+\nu \cos k_y) ~ \tau_x\otimes  \mathcal{I} + \nu \sin k_y~  \tau_y\otimes  \mathcal{I}
\end{eqnarray}
The Hermitian Hamiltonian $\mathcal{H}_0$ has both inversion and time-reversal symmetries, which lead to zero Berry curvature everywhere in Brillouin zone, except at band gap closing points. Note that the inversion operator inverts momenta $\ds{\textbf{k} \rightarrow-\textbf{k} }$ and is given by $\ds{ \tau_x\otimes \sigma_x}$. The time reversal-operator reads $\ds{ \mathcal{T} =\ \mathcal{I}\otimes  \mathcal{I} ~ \mathcal{K}}$ where $\ds{ \mathcal{K}}$ is the complex conjugation operator. The inversion symmetry remains intact only for the non-Hermitian Hamiltonian $\mathcal{H}_2 $. However, the time reversal symmetry is lost for both $\mathcal{H}_1 $ and $\mathcal{H}_2 $. It is interesting to see that only $\mathcal{H}_1 $ has combined parity-time symmetry $\ds{ \mathcal{PT} =\ \tau_x\otimes \sigma_x~ \mathcal{K}}$. Therefore energy bands for $\mathcal{H}_1 $ are real-valued unless the $\ds{ \mathcal{PT}  }$ symmetry is spontaneously broken. To this end, we note that both $\mathcal{H}_1 $ and $\mathcal{H}_2 $ have particle hole symmetry $\ds{ \mathcal{C} = \tau_z \otimes  \sigma_z ~ \mathcal{K}}$. This implies that energy eigenvalues are symmetric with respect to the zero energy line.
 \\
Let us now study their spectra. There are four energy bands since there are four sites in the unit cell. In the absence of gain and loss, the lowest and highest bands are isolated while the middle two bands touch each other at momenta $(0,0)$, $(\mp\pi,\mp\pi)$ and $(\pm\pi,\mp\pi)$. The band gap between the lowest (highest) two bands decreases with decreasing $\ds{|\omega-\nu|}$ and closes at $\ds{\omega=\nu}$, where all four band touch at the edges of the Brillouin zone. This is a signature of topological phase transition. In the non-Hermitian case, the gain and loss arrangements in the lattice plays a vital role on the reality of the spectrum. The non-Hermitian Hamiltonian $\mathcal{H}_1 $ has complex eigenvalues when $|\gamma|>\gamma_c=|\nu-\omega|$ while $\mathcal{H}_2 $ has complex eigenvalues for any nonzero value of $\gamma$. In Fig. \ref{fig2a} (a,b), we plot the imaginary parts of energy eigenvalues for  $\mathcal{H}_1 $ at $\gamma=1>\gamma_c$ and  $\mathcal{H}_2 $ at $\gamma=0.2$, respectively. Below, we will focus on $\mathcal{H}_1 $ as we are interested in finding topological states with real eigenvalues. In this case, increasing $\gamma$ at fixed $\omega$ and $\nu$  decreases the band gap between the lowest (highest) two bands and close them at the edges of the Brillouin zone at $\gamma=\gamma_c$. Beyond this value, complex eigenvalues appear around the band edges. We see that the imaginary part of the eigenvalues in $\bf{k}$-space has the shape of half-elliptic cylinder identically centered at $k_x=\mp\pi$ as can be seen from Fig. \ref{fig2a}  (a). Its length is fixed to $2\pi$ in $k_y$-direction but its semi-axis in $k_x$-direction increases with increasing $\gamma$. For large values of $\gamma$, these two shapes merge and then the spectrum becomes complex valued in the whole $\bf{k}$-space. To understand band gap opening and closing better, let us now vary $\omega$ at fixed $\nu$ and $\gamma$. The system is gapped and has real eigenvalues for large values of $\omega$. The band gap is decreased with decreasing $\omega$ and the two bands are closed at $\omega=\nu+\gamma$. In this case, exceptional points occur at the band edges. If we decrease $\omega$ further, the band gap does not reopen. Instead exceptional points (which separate the region of complex and real valued eigenvalues in $\bf{k}$-space) move in the $\bf{k}$-space until $\omega=\nu-\gamma$ at which a pair of exceptional points annihilate each other and the spectrum becomes real valued again. Generally speaking, one can not explore topological phase transition in a non-Hermitian system just by studying band gap closing and reopening. To study topological features in the system, we need to find the corresponding topological invariant. In the absence of gain and loss, the Berry curvature vanishes due to inversion and time-reversal symmetries. In the presence of gain and loss, time reversal symmetry is broken for both $\mathcal{H}_1 $ and $\mathcal{H}_2 $. However, this doesn't necessarily mean that the Chern number is quantized to be an integer multiple. Generally speaking, the standard definition of Chern number doesn't work in non-Hermitian systems. So, we need to introduce a new topological number. Below we study this issue.

\section{Complex polarization vector}

The polarization (electric dipole moment per unit volume) in a periodical potential such as solid is not uniquely defined, which leads to ambiguity for the value of the dielectric displacement field. Fortunately, not the its absolute value but the change in polarization can be measured in a real experiment. In 1993 \cite{temel0},  King-Smith and Vanderbilt formulated the polarization vector. The polarization vector is also of special importance in the theory of topological insulators \cite{temel3}. In \cite{temel1,temel2}, the authors used polarization vector to discuss topological phase in 2D Hermitian SSH system, where the Berry curvature is zero. Unfortunately, the standard polarization vector is not quantized in our non-Hermitian system, where inversion and time-reversal symmetries are broken. In fact, one can not directly apply the Hermitian polarization vector to a non-Hermitian system because of the nonorthogonal character of eigenstates in non-Hermitian systems. Our aim is to find its complex generalization.\\
Let us introduce the left-right and right-left non-Hermitian polarization vectors. The contributions from the $n$-th band read
\begin{eqnarray}\label{0yturjdu}
\textbf{P}^{LR}_n&=&\frac{e}{(2\pi)^2}\int dk_x dk_y   <\psi_n^L |~i\partial_\textbf{k}|  \psi_n^R>  \nonumber \\
\textbf{P}^{RL}_n&=&\frac{e}{(2\pi)^2}\int dk_x dk_y   <\psi_n^R |~i\partial_\textbf{k}|  \psi_n^L>
\end{eqnarray}
where the integral is taken over the 2D Brillouin zone, $n$ labels the band index, $\ds{|  \psi_n^R>  }$, $\ds{|  \psi_n^L>  }$ are the normalized right and left eigenvectors of the Hamiltonian, which form a biorthogonal basis $\ds{ <\psi_n^L|  \psi_m^R> =\delta_{nm} }$. Below, we set the charge $e=1$. Note that inversion and time-reversal symmetries are not required in the calculation of the  non-Hermitian polarization vector. We assume that the band is gapped. The non-Hermitian polarization vector is not well-defined at the EP since $\ds{ <\psi_n^L~|~  \psi_n^R>  }$ vanishes at the EP.\\
Using the condition $\ds{ <\psi_n^L|  \psi_n^R> =1}$, one can see that  $\ds{   <\psi_n^L |i \partial_\textbf{k}  \psi_n^R> =- <i\partial_\textbf{k} \psi_n^L |  \psi_n^R> = (  <\psi_n^R |i\partial_\textbf{k}  \psi_n^L>)^{\star} }$. This implies that $\ds{\textbf{P}^{LR}_n=(\textbf{P}^{RL}_n)^{\star}}$. As a result, we define the real-valued total non-Hermitian polarization vector in a two dimensional non-Hermitian gapped system as
\begin{eqnarray}\label{5774ju}
\textbf{P}=\sum_n^{occ} ~ \frac{\textbf{P}^{LR}_n+\textbf{P}^{RL}_n}{2}
\end{eqnarray}
where the summation is over the occupied states.\\
We calculate the non-Hermitian polarization vector for the lowest-lying eigenstate of the Hamiltonian $\mathcal{H}_1 $. It is quantized and equal to $\ds{  \textbf{P}_1=  (1/2,1/2)}$ for $\ds{\nu>\omega+|\gamma|}$ and $(0,0)$ for otherwise. We can define the two dimensional complex Berry (Zak) phase for the $n$-th band as $\ds{ 2{\pi} \textbf{P}_n}$. It is quantized and a topological invariant. In other words, adiabatic deformations of the Hamiltonian $\mathcal{H}_1 $ don't change it as long as the band gap is preserved. To this end, we note that the imaginary parts of $\ds{  \textbf{P}^{RL}_n   }$ cancel each other for the first (and last) two bands in pairs and their sum is also quantized. We get $\ds{\textbf{P}^{LR}_1+\textbf{P}^{LR}_2  =(1,1)}$ and $\ds{  \textbf{P}^{LR}_3+\textbf{P}^{LR}_4  =1}$ for $\ds{\nu>\omega+|\gamma|}$.  In \cite{2d6}, four different gauge invariant Berry curvatures are constructed using the combination of right and left eigenstates and the corresponding four Chern numbers are shown to be equal to each other. The above complex polarization vector is different from the complex Chern number introduced in \cite{2d6}.
\begin{figure}[t]
\includegraphics[width=4.2cm]{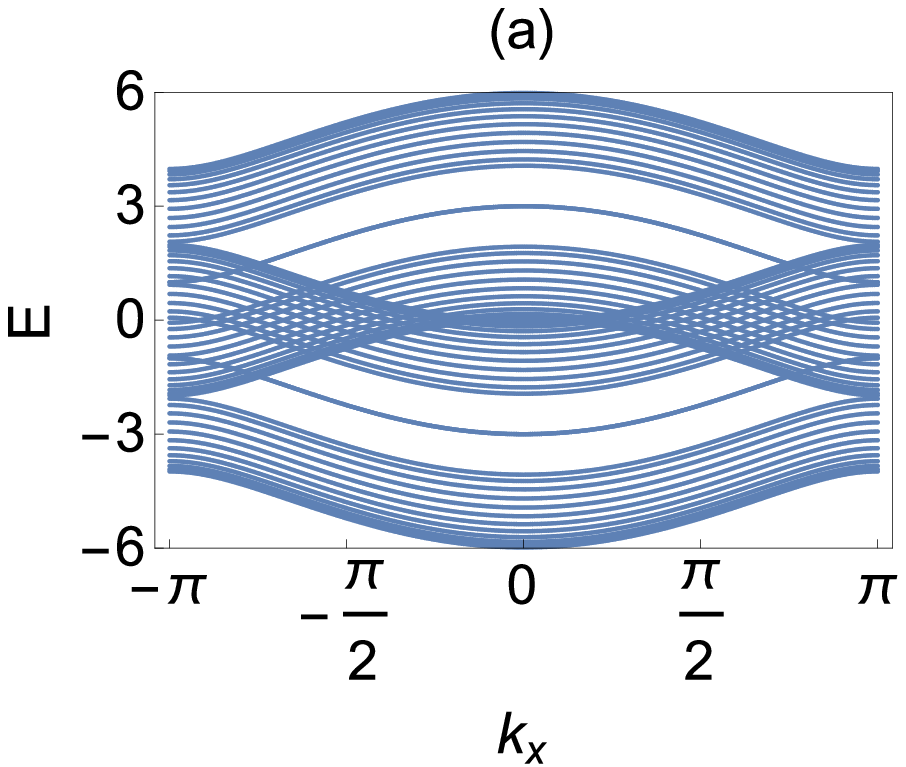}
\includegraphics[width=4.2cm]{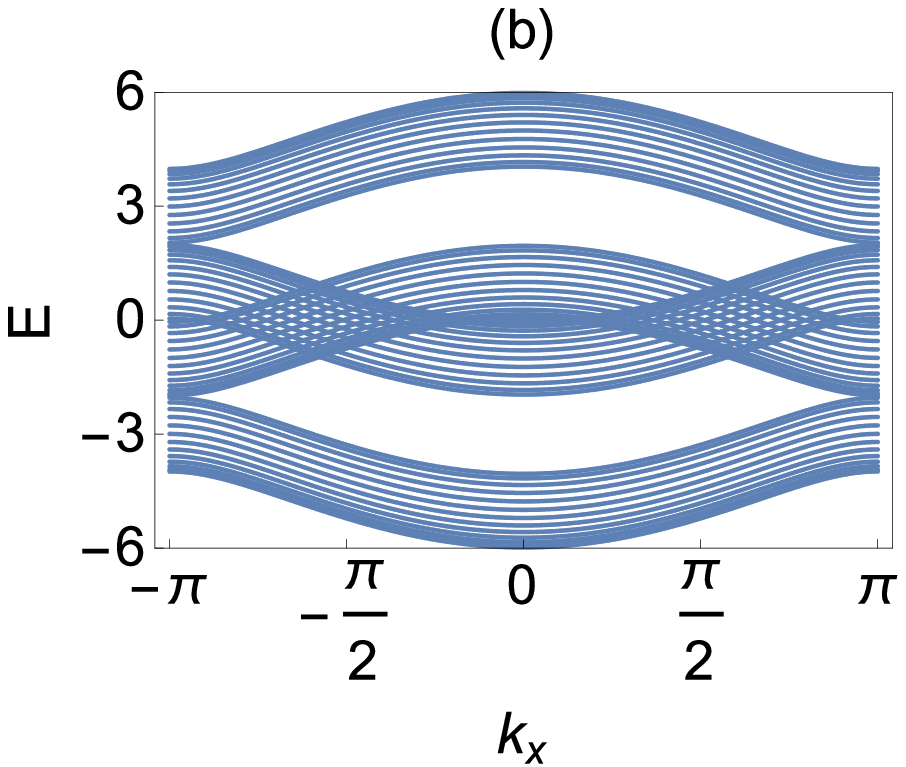}
\caption{ The spectra for the Hermitian system periodic along $x$-direction and has $N_y=52$ sites in the $y$-direction. The parameters are $\ds{\omega=1}$, $\ds{\nu=2}$ (a) and $\ds{\omega=2}$ $\ds{\nu=1}$ (b). Topological edge states appear in the band gap in (a).     }
\label{fig3a}
\end{figure}
\begin{figure}[t]
\includegraphics[width=4.2cm]{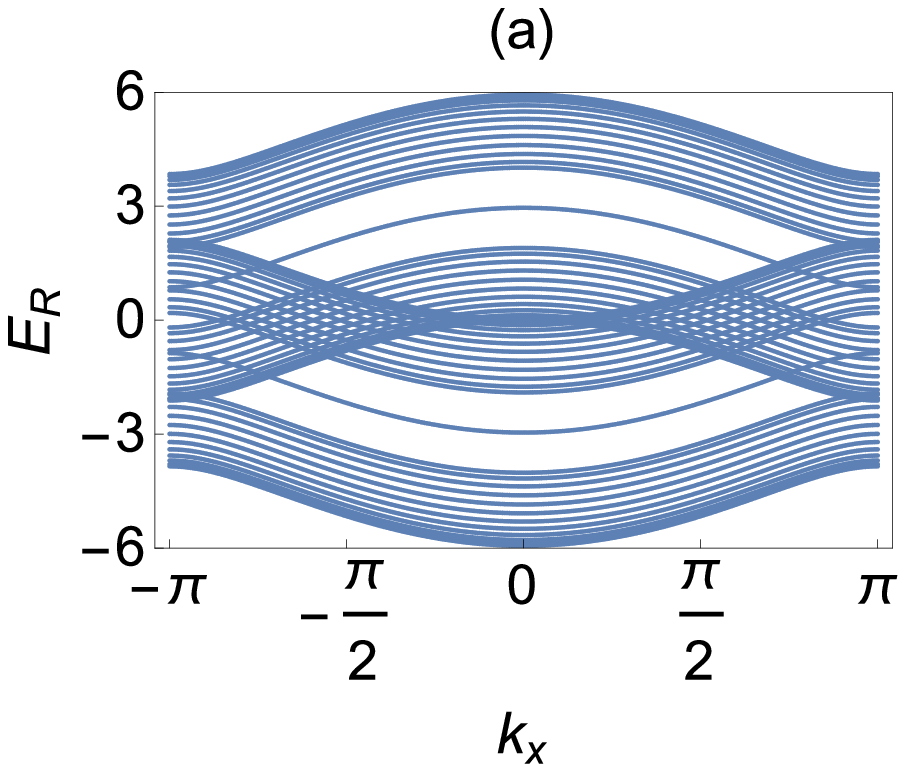}
\includegraphics[width=4.2cm]{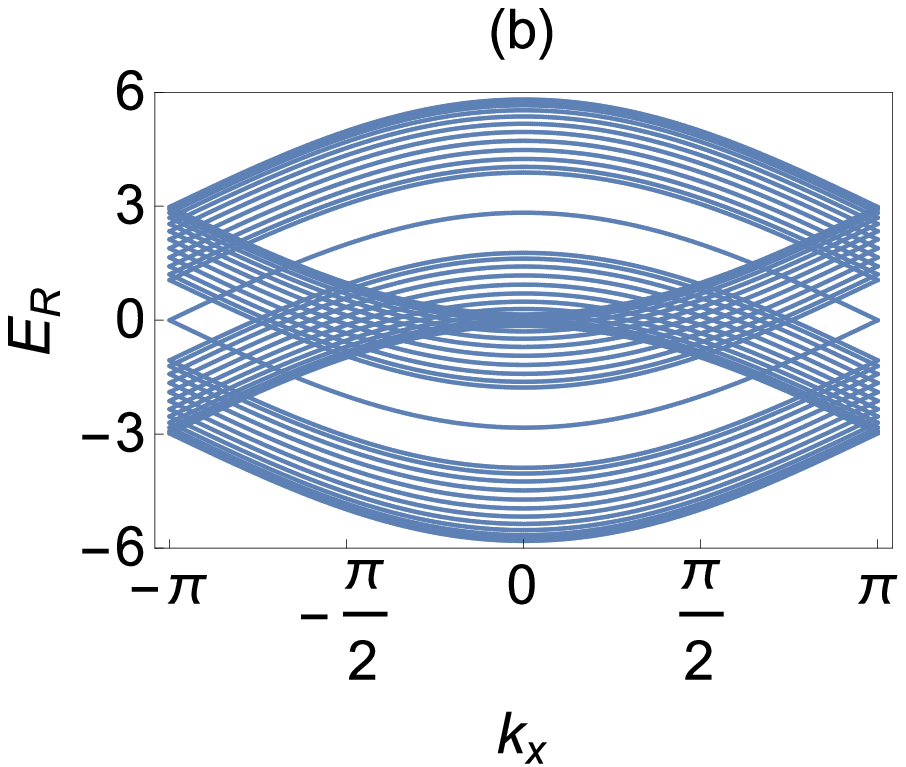}
\includegraphics[width=4.2cm]{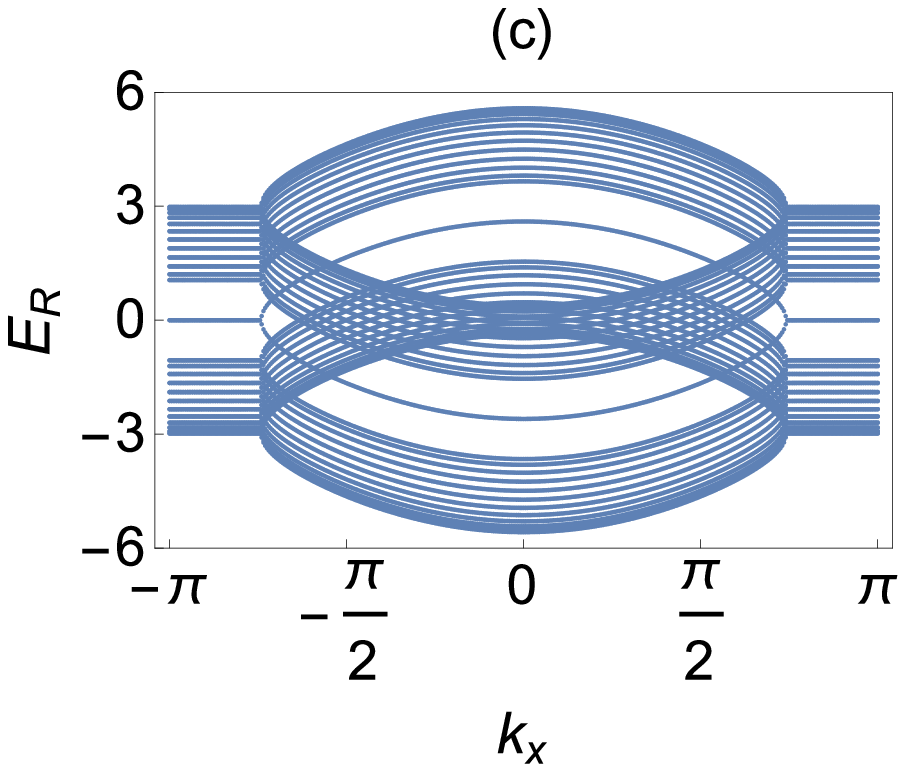}
\includegraphics[width=4.2cm]{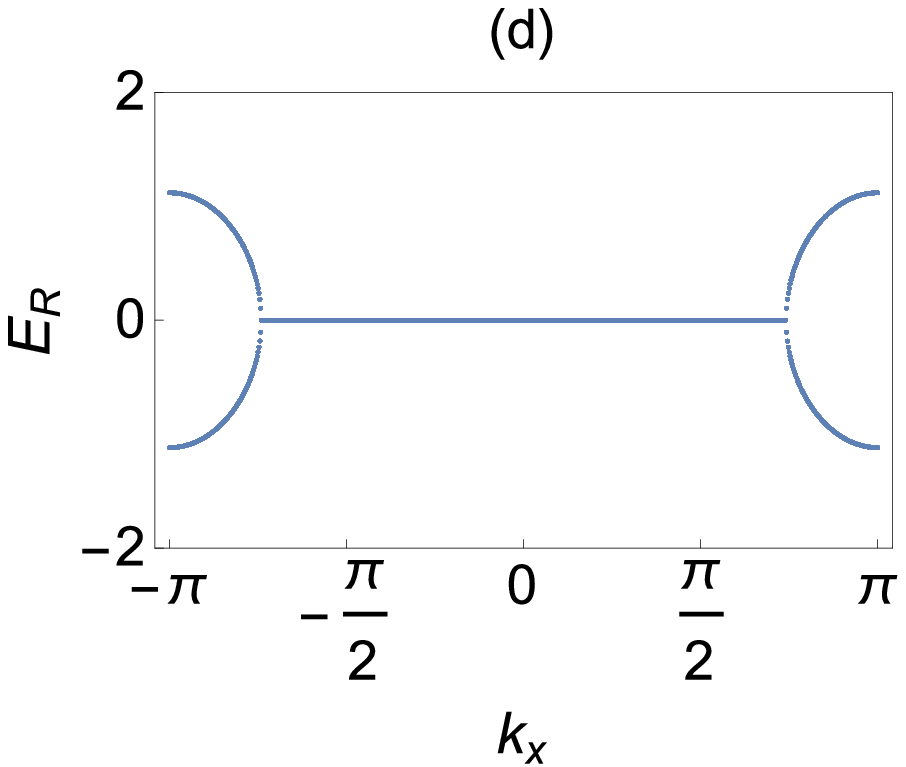}
\caption{ The non-Hermitian system $\mathcal{H}_1$ peridocal along $x$-direction and has $N_y=52$ sites in the $y$-direction for $\ds{\omega=1}$ $\ds{\nu=2}$. The real part of the energy band for $\ds{\gamma=0.5}$ (a), $\ds{\gamma=1}$ (b) , $\ds{\gamma=1.5}$ (c) and the corresponding imaginary part for $\ds{\gamma=1.5}$ (d). The spectrum becomes complex when $\ds{\gamma>1}$.   }
\label{fig4a}
\end{figure}\\
Having discussed the topological invariant, we are now in position to study topological edge states. To see them, we plot the energy spectra for a ribbon periodic along the $x$-direction and $N=52$ sites in $y$-direction. Consider first the Hermitian case $\gamma=0$. The energy spectra of the Hermitian system for the nontrivial ($\ds{\omega=1}$, $\ds{\nu=2}$) and trivial ($\ds{\omega=2}$, $\ds{\nu=1}$ ) phases are displayed in Fig.\ref{fig3a} (a,b), respectively. As can be seen from the figure, there is a distinguishing difference between them. The two energy curves within the band gaps appear in the topologically nontrivial case (a) while no such curves exist in the trivial case (b). They are in fact for the topological edge states. The edge states are doubly degenerate so there are 4 topological edge states. \\
Suppose next that $\gamma\neq0$. The complex polarization vector predicts the appearance of topological edge states. We perform numerical computation to plot the spectrum for the same nontrivial system described above but $\gamma=0.5$, $\gamma=1$ and $\gamma=1.5$ in Fig.\ref{fig4a}. One can see the topological edge states in the band gap for all three cases. These are doubly degenerate real valued topological states. Generally speaking, periodical and open boundary conditions don't predict the same critical point $\gamma_c$ for the transition from real to complex spectrum in a non-Hermitian topological system. Fortunately, this is not the case in our case. Therefore the spectra are real in Fig.\ref{fig4a} (a,b) while it is complex around band edges in Fig.\ref{fig4a} (c).  One can also see that the spectra are symmetric with respect to $E=0$ line. This is because of the particle-hole symmetry of the system.  \\
There is a fundamental difference between the Hermitian and non-Hermitian systems. In the Hermitian SSH, topological edge states disappear whenever the band gap is zero, which occurs at $\nu=\omega$. However, this is not the case in the non-Hermitian system. At $\gamma=\gamma_c=|\nu-\omega|$, the gap vanishes at $k_x=\mp \pi$ as can be seen from Fig.\ref{fig4a} (b). In other words, exceptional points occur at $\gamma=\gamma_c$ and $k_x=\mp \pi$. Topological edge states still exist even if $\gamma>\gamma_c$ as can be seen from Fig.\ref{fig4a}  (c). They have purely imaginary eigenvalues in between $k_x\in[\mp\pi,{\mp}k_c]$, where $k_c$ increases with $\gamma$. For very large values of the non-Hermitian strength, $\gamma>\nu+\omega$, the eigenvalues become purely imaginary in the whole $k_x$-space. The complex Zak phase based on the non-Hermitian polarization vector predicts the appearance of topological edge states correctly as long as $\gamma\leq\gamma_c$. However, it does not work if  $\gamma>\gamma_c$ since the band gap is closed in this case. Note that the non-Hermitian polarization vector works only for gapped Hamiltonian. To this end, we say that topological edge states appear also for the other Hamiltonian $\mathcal{H}_2 $. But they have purely imaginary eigenvalues. This means that it can be used as a prototype for topological laser in 2D.\\
To sum up, we have studied a non-Hermitian square lattice, which can be experimentally realized in photonics using waveguides. The corresponding Hamiltonian is a complex extension of the 2D SSH Hamiltonian. We have shown that topological edge states with real eigenvalues appear in the system as long as the non-Hermitian strength is below than a critical value. We have introduced non-Hermitian polarization vector, which predicts the topological edge states when the spectrum is real. We have shown that topological edge states survive even when the Hamiltonian is gapless.


\begin{thebibliography}{0}
\bibitem{ghatakdas} Ananya Ghatak, and Tanmoy Das, J. Phys. Condens. Matter {\bf 31}, 263001 (2019).
\bibitem{1d5} V. M. Martinez Alvarez, J. E. Barrios Vargas, M. Berdakin, L. E. F. Foa Torres, Eur. Phys. J. Special Topics {\bf 227}, 1295 (2018).
\bibitem{kato} T. Kato, Perturbation Theory for Linear Operators (Springer-Verlag, Berlin, 1966).
\bibitem{1d1} L. Jin, Phys. Rev. A {\bf 96}, 032103 (2017).
\bibitem{1d2} Chuanhao Yin, Hui Jiang, Linhu Li, Rong Lu, and Shu Chen, Phys. Rev. A {\bf 97}, 052115 (2018).
\bibitem{1d3} C. Yuce, Phys. Rev. A {\bf 98}, 012111 (2018).
\bibitem{1d3ekl} C. Yuce and Z. Oztas, Sci. Rep. {\bf 8}, 17416 (2018).
\bibitem{1d4}  Samit Kumar Gupta, et. al., arXiv:1803.00794 (2018).
\bibitem{1d6} L. Jin, P. Wang, Z. Song, Sci. Rep. {\bf 7}, 5903 (2017).
\bibitem{1d7} Kun Ding, Z. Q. Zhang, and C. T. Chan, Phys. Rev. B {\bf 92}, 235310 (2015).
\bibitem{1d8} C. W. Ling, Ka Hei Choi, T. C. Mok, Z. Q. Zhang, Kin Hung Fung, Sci. Rep. {\bf 6}, 38049 (2016).
\bibitem{1d9} Li-Jun Lang, You Wang, Hailong Wang, and Y. D. Chong, Phys. Rev. B {\bf 98}, 094307 (2018).
\bibitem{1d10} Mingsen Pan, Han Zhao, Pei Miao, Stefano Longhi and Liang Feng, Nat. Commun. {\bf 9}, 1308 (2018).
\bibitem{1d11} X. Z. Zhang, Z. Song, Phys. Rev. A {\bf 99}, 012113 (2019).
\bibitem{1d12} B. X. Wang, C. Y. Zhao, Phys. Rev. B  {\bf 98}, 165435 (2018).
\bibitem{1d13} Simon Lieu, Phys. Rev. B {\bf 97}, 045106 (2018).
\bibitem{1d14} Ze-Zhong Li, Xue-Si Li, Lian-Lian Zhang, Wei-Jiang Gong, arXiv:1901.10688 (2019).
\bibitem{1d15} Bikashkali Midya and Liang Feng, Phys. Rev. A {\bf 98}, 043838 (2018).
\bibitem{winding1} Chuanhao Yin, Hui Jiang, Linhu Li, Rong Lu, and Shu Chen, Phys. Rev. A {\bf 97}, 052115 (2018).
 \bibitem{aah1} C. Yuce, Phys. Lett. A {\bf 379}, 1213 (2015).
\bibitem{kita1} C. Yuce Phys. Rev. A {\bf 93}, 062130 (2016).
\bibitem{kita2} C. Li, X. Z. Zhang, G. Zhang, and Z. Song, Phys. Rev. B {\bf 97}, 115436 (2018).
\bibitem{kita3} Marcel Klett, Holger Cartarius, Dennis Dast, Jorg Main, and Gunter Wunner, Phys. Rev. A {\bf 95}, 053626 (2017).
\bibitem{kita4} Kohei Kawabata, Yuto Ashida, Hosho Katsura, and Masahito Ueda, Phys. Rev. B {\bf 98}, 085116 (2018).
\bibitem{kita5} Henri Menke and Moritz M. Hirschmann, Phys. Rev. B {\bf 95}, 174506 (2017).
\bibitem{bulkboun01} V. M. Martinez Alvarez, J. E. Barrios Vargas, L. E. F. Foa Torres, Phys. Rev. B {\bf 97}, 121401(R) (2018).
\bibitem{bulkboun02} Flore K. Kunst, Elisabet Edvardsson, Jan Carl Budich, Emil J. Bergholtz, Phys. Rev. Lett. {\bf 121}, 026808 (2018). 
\bibitem{bulkboun03} Shunyu Yao, Zhong Wang, Phys. Rev. Lett. {\bf 121}, 086803 (2018).
\bibitem{bulkboun04} Shunyu Yao, Fei Song, Zhong Wang, Phys. Rev. Lett.  {\bf 121}, 136802 (2018).
\bibitem{bulkboun05} C. Yuce, Phys. Rev. A {\bf 97}, 042118 (2018).
\bibitem{bulkboun06} Daniel Leykam, Konstantin Y. Bliokh, Chunli Huang, Y. D. Chong, and Franco Nori, Phys. Rev. Lett. {\bf 118}, 040401 (2017).
\bibitem{bulkboun07} L. Jin, Z. Song, Phys. Rev. B {\bf 99}, 081103(R) (2019).
\bibitem{bulkboun08da} Kazuki Yokomizo, Shuichi Murakami, arXiv:1902.10958 (2019).
\bibitem{takata} Kenta Takata and Masaya Notomi, Phys. Rev. Lett. {\bf 121}, 213902 (2018).
\bibitem{pseudo} C. Yuce, Phys. Let. A {\bf 383}, 248 (2019).
\bibitem{yucepump} C. Yuce, Phys. Rev. A  {\bf 99}, 032109 (2019).
\bibitem{merced} G. Harari, et al. Topological insulator laser: theory. Science {\bf 359}, eaar4003 (2018).
\bibitem{merced2} Yasutomo Ota, Ryota Katsumi, Katsuyuki Watanabe, Satoshi Iwamoto and Yasuhiko Arakawa, Communications Physics  {\bf 1}, 86 (2018). 
\bibitem{sondeney1} S. Weiman, et. al., Nat. Mater. {\bf 16}, 433 ( 2017).
\bibitem{2d1} Z Oztas, C Yuce, Phys. Rev. A  {\bf 98}, 042104 (2018).
\bibitem{2d2} Mark Kremer, Tobias Biesenthal, Matthias Heinrich, Ronny Thomale, Alexander Szameit, Nat. Commun. {\bf 10}, 435 (2019).
\bibitem{2d3} Xue-Yi Zhu, Samit Kumar Gupta, Xiao-Chen Sun, Cheng He, Gui-Xin Li, Jian-Hua Jiang, Xiao-Ping Liu, Ming-Hui Lu, and Yan-Feng Chen, Opt. Express {\bf 26}, 24307 (2018).
\bibitem{2d4} Kohei Kawabata, Ken Shiozaki, Masahito Ueda, Phys. Rev. B {\bf 98}, 165148 (2018).
\bibitem{2d5} R. Wang, X. Z. Zhang, and Z. Song, Phys. Rev. A {\bf 98}, 042120 (2018).
\bibitem{2d6} Huitao Shen, Bo Zhen, and Liang Fu, Phys. Rev. Lett. {\bf 120}, 146402 (2018).
\bibitem{2d7} Z. Ozcakmakli Turker, C. Yuce, Phys. Rev. A {\bf 99}, 022127 (2019).
\bibitem{2d8} Huaiqiang Wang, Jiawei Ruan, and Haijun Zhang, Phys. Rev. B {\bf 99}, 075130 (2019).
\bibitem{2d9} Jan Carl Budich, Johan Carlstrom, Flore K. Kunst, and Emil J. Bergholtz, Phys. Rev. B {\bf 99}, 041406(R) (2019).
\bibitem{2d10} Kristof Moors, Alexander A. Zyuzin, Alexander Yu. Zyuzin, Rakesh P. Tiwari, and Thomas L. Schmidt, Phys. Rev. B {\bf 99}, 041116(R) (2019).
\bibitem{2d11} Kaifa Luo, Jiajin Feng, Y. X. Zhao, Rui Yu, arXiv:1810.09231 (2018).
\bibitem{2d12} Tsuneya Yoshida, Robert Peters, Norio Kawakami, Yasuhiro Hatsugai, Phys. Rev. B {\bf 99}, 121101 (2019).
\bibitem{2d13} Zhesen Yang and Jiangping Hu, Phys. Rev. B  {\bf 99}, 081102(R) (2019).
\bibitem{temel0} R. D. King-Smith and David Vanderbilt, Phys. Rev. B {\bf 47}, 1651(R) (1993).
\bibitem{temel3} Wladimir A. Benalcazar, B. Andrei Bernevig, Taylor L. Hughes, Science {\bf 357}, 61 (2017).
\bibitem{temel1} F. Liu, and K. Wakabayashi, Phys. Rev. Lett. {\bf 118}, 076803 (2017).
\bibitem{temel2} F. Liu, H. Y. Deng, and K. Wakabayashi. Phys. Rev. B {\bf 97}, 035442 (2018).

\end{thebibliography}
\end{document}